\documentclass[12pt]{article}
\usepackage{amssymb,amsmath,epsfig}
\allowdisplaybreaks
\begin{document}

\title{\bf Hydrodynamics of Gaseous System in Massive Brans-Dicke Gravity}

\author{M. Sharif $^1$ \thanks{msharif.math@pu.edu.pk} and Rubab
Manzoor $^{1,2}$
\thanks{rubab.manzoor@umt.edu.pk}\\
$^1$ Department of Mathematics, University of the Punjab,\\
Quaid-e-Azam Campus, Lahore-54590, Pakistan.
\\$^2$ Department of Mathematics,\\ University of Management and
Technology,\\
Johar Town Campus, Lahore-54782, Pakistan.}
\date{}
\maketitle

\begin{abstract}
This paper explores hydrodynamics and hydrostatic of a star in
post-Newtonian approximation of massive Bran-Dicke gravity. We study
approximated solution of the field equations upto $O(c^{-4})$ and
generalize Euler equation of motion. We then formulate equations
governing hydrodynamics, stability and instability of the system.
Finally, we discuss spherically symmetric stars for a specific
barptropic case like dust, cosmic string and domain wall in this
scenario.
\end{abstract}
{\bf Keywords:} Hydrodynamics; Brans-Dicke Theory; Newtonian and
post-Newtonian regimes.\\
{\bf PACS:} 04.50.Kd; 04.25.Nx.

\section{Introduction}

The mysteries of the universe always remain interesting issue for
the physicists. The combination of two theories, i.e., theory of
fluid dynamics and theory of gravity play a major role in the study
of the universe. It describes universe phenomena  in three contexts:
(i) dynamics of celestial objects in weak-field regimes (like
dynamics of solar system), (ii) dynamics of strong-field regimes
such as Supernovae, formation of black hole, superluminal jets,
binaries pulsars, neutron stars, X-rays and Gamma-rays burst etc and
(iii) cosmological issues (origin, different eras and eventual final
fate of the universe) \cite{1}.

In this context, the concept of weak-field approximation (Newtonian
and post-Newtonian (pN) approximations) of relativistic
hydrodynamics has taken considerable importance in many regards. It
provides a platform where we can evaluate ranges of deviation and
level of consistency between relativistic theory of gravity and
Newton's gravity. The analysis of phenomena in strong-field regimes
are highly complicated and non-linear, so in order to obtain results
of physical interest, weak-field approximation schemes are used as
an effective tool \cite{2*}. Chandrasekhar \cite{2} combined theory
of hydrodynamics and general relativity (GR) at pN limits to remove
difficulties occurring in the analysis of large scale structures. He
derived solutions of the field equations in pN correction (order of
$c^{-4}$) of GR which contain potential and superpotential functions
to represent masses of gaseous or stellar structure. He then used
these solutions to derive a set of generalized Newtonian
hydrodynamics equations and discussed complicated issues of gaseous
structures (like radial as well as non-radial oscillation of stars,
rotating homogenous masses and stability of gaseous masses under
non-radial as well as radial oscillation) in a approximated region
in which strong field interactions (gravitational radiation) play no
role. In this way, he represented a mechanism (in weak-field
approximation) which can analyze different gaseous or stellar
systems in GR \cite{3}. After his work, many researchers \cite{4}
explored different astrophysical systems in weak-field approximation
of GR and obtained many interesting results.

The modified theories of gravity are considered the most fascinated
approaches to resolve mysteries of the present universe ( the dark
energy and dark matter). Brans-Dicke (BD) gravity \cite{5} is an
attractive example of modified gravity in which gravity is mediated
by a massless scalar field $\phi$ and curvature. It contains a
coupling constant $\omega_{BD}$ which serves as a tuneable parameter
and can adjust results according to the requirement. This theory has
provided convenient solutions of many cosmic issues but unable to
explain ``graceful exist" problem of old inflationary according to
observational surveys. The inflationary model defined by BD gravity
is valid for specific ranges of coupling parameter
($\omega_{BD}\leq25$) \cite{5a} which is not consistent with
observational limits \cite{5b}.  Moreover, the BD gravity probes
strong field test (cosmic issues) for negative and low values of
$\omega_{BD}$ \cite{7} but satisfies weak field test for high and
positive values of $\omega_{BD}$ \cite{6}. Thus, the weak field
evaluations are not consistent with the results of strong field.

In order to resolve this issue, a scalar potential function
$V(\phi)$ (massive function) is introduced in BD gravity which leads
to massive Brans-Dicke (MBD) gravity \cite{6a}. This new theory not
only solves cosmic issue (old inflation) but also provides a
consistency between the results obtained for weak-field as well as
cosmic scale \cite{8}. There is a large body of literature which
describes dynamics of the universe in modified gravity \cite{9}. In
this context, Nutku \cite{10} explained fluid hydrodynamics in this
gravity. He represented approximated solutions of BD equations in
complete pN limits (order of $c^{-4}$) involving potential functions
(representing masses of the celestial objects) and explored
stability of spherically symmetric gaseous system. Olmo \cite{11}
evaluated complete pN limits of MBD field equations solutions but he
converted only lowest-order (order of $c^{-2})$ approximation of
solutions in terms of potential functions to discuss $f(R)$ gravity
as a special case of scalar-tensor gravity.

In order to provide a suffice platform for the analysis of gaseous
systems in accelerated expanding universe, we represent
hydrodynamics of fluid in complete pN limits of MBD gravity. For
this purpose, in section \textbf{1}, we obtain complete pN
approximation (order of $c^{-4}$) of MBD field solutions in terms of
potential and superpotential functions. In section \textbf{3} we
formulate equations governing hydrodynamics of the fluid in pN
regimes. Section \textbf{4}, evaluates conditions that govern
stability and instability. Section \textbf{5} explores spherically
symmetric barotropic stars in MBD gravity. Finally, section
\textbf{6} summarizes the results.

\section{Massive Brans-Dicke Gravity in Post-\\Newtonian Limits}

The action of MBD gravity with ($\kappa^{2}=\frac{8\pi G}{c^{2}}$)
\cite{8} is
\begin{equation}\label{1}
S=\frac{1}{2\kappa^{2}}\int d^{4}x\sqrt{-g} [\phi
R-\frac{\omega_{BD}}{\phi}\nabla^{\alpha}{\phi}\nabla_{\alpha}{\phi}-V(\phi)]
+L_{m}[g,\psi],
\end{equation}
where $L_{m}[g,\psi']$ represents matter action which depends upon
metric as well as matter field $\psi'$. Variation of the above
action with respect to $g_{\alpha\beta}$ and $\phi$ provides
respective MBD equations
\begin{eqnarray}\nonumber
G_{\alpha\beta}&=&\frac{\kappa^{2}}{\phi}T_{\alpha\beta}+[\phi_{,\alpha;\beta}
-g_{\alpha\beta}\Box\phi]+\frac{\omega_{BD}}{\phi}[\phi_{,\alpha}\phi_{,\beta}
-\frac{1}{2}g_{\alpha\beta}\phi_{,\mu}\phi^{,\mu}]
-\frac{V(\phi)}{2}g_{\alpha\beta},\\\label{2}
\\\label{3}
\Box\phi&=&\frac{\kappa^{2}T}{3+2\omega_{BD}}
+\frac{1}{3+2\omega_{BD}}[\phi\frac{dV(\phi)}{d\phi}-2V(\phi)].
\end{eqnarray}
Here $T_{\alpha\beta}$ is the energy-momentum tensor of matter,
$T=g^{\alpha\beta}T_{\alpha\beta}$ and $\Box$ is the d'Alembertian
operator. Equations (\ref {2}) and (\ref{3}) describe the MBD field
equations and evolution equation for the scalar field, respectively.
We take matter contribution in the form of perfect fluid which is
compatible with pN approximation
\begin{equation}\label{4}
T_{\alpha\beta}=[\rho
c^2(1+\frac{\pi}{c^2})+p]u_{\alpha}u_{\beta}-pg_{\alpha\beta},
\end{equation}
where $\rho,~\rho\pi,~p$ and $u_{\mu}$ represent matter density,
thermodynamics internal energy, pressure and four velocity,
respectively.

In order to calculate some approximated MBD solutions in pN regime,
the following Taylor expansions are assumed \cite{11}
\begin{eqnarray}\nonumber
g_{\alpha\beta}&\approx&\eta_{\alpha\beta}+h_{\alpha\beta},\quad
\phi\approx\phi_{0}+\varphi^{(2)}+\varphi^{(4)},\\\nonumber
V(\phi)&\approx& V_{0}+\varphi V'_{0}+\varphi^{2}V''_{0}/2+....
\end{eqnarray}
Here $\eta_{\alpha\beta}$ is the Minkoski metric (representing
non-dynamical background), $h_{\alpha\beta}$ shows the perturbation
tensor which describes deviation of $g_{\alpha\beta}$ from
$\eta_{\alpha\beta}$, the term $\phi_{0}$ indicates asymptotic
cosmic function which slowly varies  with respect to cosmic time
$t_{0},~\varphi(t,x)$ represents local deviation of scalar field
from $\phi_{0}$ with superscripts ($(2)$ and $(4)$) indicating order
of approximation $(c^{-2})$ as well as $(c^{-4})$ and
$V_{0}=V(\phi_{0})$ shows the value of potential function at time
$t_{0}$. The pN limits of MBD theory are obtained through the gauge
condition
\begin{equation}\nonumber
h^{\alpha}_{k,\alpha}-\frac{1}{2}h^{\alpha}_{\alpha,k}
-\frac{\partial_{k}\varphi}{c^{2}\phi_{0}}=0,
\end{equation}
and the lowest-order $(O(2))$ pN approximated solutions describing
the potential of the compact object are given by
\begin{eqnarray}\label{4}
g_{00}&\approx& 1-h^{(2)}_{00}=1-\frac{2U}{c^{2}}
+\frac{\Lambda_{BD}r^{2}}{3c^{2}},\\\label{4a} g_{ij}&\approx&
-[1+h^{(2)}_{ij}]\delta_{ij}=[-1-\frac{2\gamma_{BD}
U}{c^{2}}-\frac{\Lambda_{BD}r^{2}}{3c^{2}}]\delta_{ij},\\\label{4b}
\frac{\varphi}{\phi_{0}}&\approx&\frac{-2U}{c^{2}}
\left[\frac{e^{-m_{0}r}}{3+2\omega_{BD}+e^{-m_{0}r}}\right],
\end{eqnarray}
where  $i,j=1,2,3$ and $U$ is the gravitational potential determined
by Poisson's equation
\begin{equation}\label{4aa}
\nabla^{2}U=-4\Pi\rho G_{eff}
\end{equation}
with
\begin{equation}\nonumber
G_{eff}=\frac{\kappa^{2}}{8\pi\phi_{0}}\left(1
+\frac{e^{-m_{0}r}}{3+2\omega_{BD}}\right),\quad
m_{0}=\left(\frac{\phi_{0}V''_{0}-V'_{0}}{3+2\omega_{BD}}\right)^{1/2}.
\end{equation}
The term  $\Lambda_{BD}=\frac{V_{0}}{2\phi_{0}}$ shows the
cosmological constant, $\gamma_{BD}$ is the parameterized pN
parameter given by
\begin{equation}\nonumber
\gamma_{BD}=\frac{3+2\omega_{BD}-e^{-m_{0}r}}{3+2\omega_{BD}+e^{-m_{0}r}}.
\end{equation}

To discuss characteristics of celestial fluid in complete pN limits
($O(4)$), we evaluate complete pN solutions of MBD equation in terms
of potential and super-potential functions \cite{2, 10}. For this
purpose, we calculate values of $g_{00} \sim O(4),~g_{0i} \sim O(3)$
in terms of potential and super-potential functions. Consider the
following metric coefficients
\begin{equation}\label{5}
g_{00}=1+h_{00}, \quad g_{0i}=h_{0i}, \quad
g_{ij}=-\delta_{ij}+h_{ij},
\end{equation}
where
\begin{equation}\nonumber
h_{00}=h^{(2)}_{00}+O(4),\quad h_{0i}=O(3), \quad
h_{ij}=h^{(2)}_{ij}.
\end{equation}
The components of four-velocity and energy-momentum tensor can be
found by using Eq.(\ref{5}) given in appendix \textbf{A}. Equation
(\ref{2}) can also be written as
\begin{eqnarray}\label{5'}
R_{\alpha\beta}=\frac{\kappa^{2}}{\phi}(T_{\alpha\beta}-\frac{1}{2}g_{\alpha\beta}T)
+\frac{\omega_{BD}}{\phi^{2}}[\phi_{,\alpha}\phi_{,\beta}]+\frac{1}{\phi}[\phi_{,\alpha;\beta}]
-\frac{g_{\alpha\beta}}{2\phi}[\Box\phi+V(\phi)].
\end{eqnarray}
The $(0, i)$ and (0,0) components of above equation are given by
\cite{11}
\begin{eqnarray}\label{5a}
&&\frac{-1}{2}\nabla^{2}h^{(3)}_{0i}-\frac{1}{4}h^{(2)}_{00,0j}+8\Pi
G_{eff}\frac{3+2\omega_{BD}}{3+2\omega_{BD}+e^{-m_{0}r}}\rho
v_{i}=0,\\\nonumber
&&\frac{-1}{2}\nabla^{2}\left[h^{(4)}_{00}+\frac{(h^{(2)}_{00})^{2}}{2}
+\frac{1}{2}(\frac{\varphi^{(2)}}{\phi_{0}})^2\right]
=\frac{\kappa^{2}\rho}{2\phi_{0}}\left[c^{2}+\pi+2v^{2}+h^{(2)}_{[ij]}
-\frac{\varphi^{(2)}}{\phi_{0}}\right.\\\label{5a'}
&&\left.+\frac{3p}{\rho}\right]-\frac{1}{2\phi_{0}}\left[V_{0}(1+h^{(2)}_{[ij]}
-\frac{\varphi^{(2)}}{\phi_{0}})+\varphi^{(2)}V'_{0}\right].
\end{eqnarray}
Here the effect of $\phi_{0}$ is considered almost constant, hence
the contributions due to $\dot{\phi_{0}}$ and $\ddot{\phi_{0}}$ are
neglected.

In order to solve Eq.(\ref{5a}) for $h^{(3)}_{0i}$, we assume that
the second and third term of the equation can be expressed in terms
of the potential functions $\chi$ as well as $U_{i}$ represented by
the Poisson equations as
\begin{eqnarray}\label{5b}
&&\nabla^{2}\chi=h^{(2)}_{00}=\frac{1}{c^{2}}(-2U+\frac{\Lambda_{BD}r^{2}}{3}),\\\label{5c}
&&\nabla^{2}\left(\frac{3+2\omega_{BD}}{3+2\omega_{BD}+e^{-m_{0}r}}\right)U_{i}=-4\Pi
G_{eff}\frac{3+2\omega_{BD}\rho v_{i}}{3+2\omega_{BD}+e^{-m_{0}r}},
\end{eqnarray}
where $U_{i}=-4\Pi G \rho v_{i}$ \cite{2}. Equations (\ref{5a}),
(\ref{5b}) and (\ref{5c}) give
\begin{equation}\nonumber
\frac{-1}{2}\nabla^{2}h^{(3)}_{0i}-\frac{1}{4}\nabla^{2}\frac{\partial^{2}\chi}{\partial
t\partial x_{i}}
-2\nabla^{2}\left(\frac{3+2\omega_{BD}}{3+2\omega_{BD}+e^{-m_{0}r}}\right)U_{i}=0,
\end{equation}
which provides
\begin{eqnarray}\label{6a}
h^{(3)}_{0i}=\frac{1}{c^{3}}\left(4U_{i}\left(\frac{3+2\omega_{BD}}{3+2\omega_{BD}+e^{-m_{0}r}}\right)
-\frac{1}{2}\frac{\partial^{2}\chi}{\partial t\partial
x_{i}}\right).
\end{eqnarray}
Now we verify that this solution (\ref{6a}) is consistent with gauge
condition as
\begin{eqnarray}\nonumber
&&h^{\alpha}_{0,\alpha}-\frac{1}{2}h^{\alpha}_{\alpha,0}
-\frac{\partial_{0}\varphi}{c^{2}\phi_{0}}=
\frac{4}{c^{3}}\left[\frac{3+2\omega_{BD}}{3+2\omega_{BD}+e^{-m{0}r}}\right]\\\label{6a'}
&&\left[\frac{\partial U}{\partial
t}+\frac{3+2\omega_{BD}+e^{-m_{0}r}}{3+2\omega_{BD}}\frac{\partial}{\partial
x_{i}}\left(\frac{3+2\omega_{BD}}{3+2\omega_{BD}+e^{-m_{0}r}}\right)U_{i}\right].
\end{eqnarray}
From Eqs.(\ref{4aa}), (\ref{5c}) and Newtonian equation of
continuity, we have
\begin{eqnarray}\nonumber
&&\nabla^{2}\left[\frac{\partial U}{\partial
t}+\frac{3+2\omega_{BD}+e^{-m_{0}r}}{3+2\omega_{BD}}\frac{\partial}{\partial
x_{i}}\left(\frac{3+2\omega_{BD}}{3+2\omega_{BD}+e^{-m_{0}r}}\right)U_{i}\right]\\\nonumber
&&=\frac{-4\pi G_{eff}}{c^{3}}\left[\frac{\partial \rho}{\partial
t}+\frac{\partial}{\partial x_{i}}(\rho v_{i})\right]=0,
\end{eqnarray}
which implies
\begin{equation}\label{6b'}
\left[\frac{\partial U}{\partial
t}+\frac{3+2\omega_{BD}+e^{-m_{0}r}}{3+2\omega_{BD}}\frac{\partial}{\partial
x_{i}}\left(\frac{3+2\omega_{BD}}{3+2\omega_{BD}
+e^{-m_{0}r}}\right)U_{i}\right]=0.
\end{equation}
Equations (\ref{6a'}) and (\ref{6b'}) show the consistency of the
solution with gauge condition.

Now we evaluate  $h^{(4)}_{00}$ from Eq.(\ref{5a'}). For this
purpose, we consider that the left hand side of Eq.(\ref{5a'}) can
be defined in terms of potential function $\psi$ and super-potential
$\Phi$ given by the following Poisson equations \cite{2, 10}
\begin{eqnarray}\label{6'}
\nabla^{2}\psi&=&-\frac{1}{4\phi_{0}}\left[V_{0}(1+h^{(2)}_{[ij]}
-\frac{\varphi^{(2)}}{\phi_{0}})+\varphi^{(2)}V'_{0}\right],\\\label{7}
\nabla^{2}\Phi'&=&-4\Pi G_{eff}\rho\sigma,\quad \Phi'=\Phi+2\psi,
\end{eqnarray}
where
\begin{equation}\nonumber
\sigma=\frac{3+2\omega_{BD}}{3+2\omega_{BD}+e^{-m_{0}r}}
\left[\pi+2v^{2}-h^{(2)}_{00}
-\frac{\varphi^{(2)}}{\phi_{0}}+\frac{3p}{\rho}\right].
\end{equation}
Equations (\ref{4aa}), (\ref{5a'}), (\ref{6'}) and (\ref{7}) give
\begin{eqnarray}\nonumber
\nabla^{2}\left[h^{(4)}_{00}+\frac{(h^{(2)}_{00})^{2}}{2}
+\frac{1}{2}(\frac{\varphi^{(2)}}{\phi_{0}})^2+\frac{2U}{c^{4}}
(\frac{3+2\omega_{BD}}{3+2\omega_{BD}-e^{-m_{0}r}})+2\Phi+2\psi\right]=0
\end{eqnarray}
which implies
\begin{eqnarray}\label{8}
h^{(4)}_{00}=-\frac{(h^{(2)}_{00})^{2}}{2}
-\frac{1}{2}(\frac{\varphi^{(2)}}{\phi_{0}})^2-\frac{2U}{c^{4}}
(\frac{3+2\omega_{BD}}{3+2\omega_{BD}+e^{-m_{0}r}})-2\Phi-2\psi.
\end{eqnarray}
The complete pN corrections of MBD gravity are given in
Eqs.(\ref{4})-(\ref{4b}), (\ref{6a}) and (\ref{8}). Notice that all
these solutions are consistent with GR in the limits
$V_{0}\ll1,~\mid\frac{\varphi^{(2)}}{\phi_{0}}\mid \ll1$ and
$\omega_{BD}\rightarrow \infty$.

\section{Hydrodynamics of Fluid in Massive Brans-Dicke Gravity}

Hydrodynamics of fluid in any theory is governed by three laws given
by
\begin{itemize}
\item Law of conservation of mass,
\item Law of conservation of momentum,
\item Law of conservation of energy.
\end{itemize}
In Newtonian theory, these laws are evaluated through equation of
continuity and Euler equation of motion. In relativistic
hydrodynamics, these law are determined through relativistic
(generalized) form of equation of continuity and Euler equation of
motion which are obtained from the following identity \cite{2}
\begin{equation}\label{9}
T^{\alpha\beta}_{;\beta}=0.
\end{equation}

\subsection{Generalized Equation of Continuity and Euler\\ Equation of motion}

According to pN corrections of MBD theory, we have
\begin{eqnarray}\nonumber
g_{00}&\approx&1-\frac{(h^{(2)}_{00})^{2}}{2}
-\frac{1}{2}(\frac{\varphi^{(2)}}{\phi_{0}})^2-\frac{2U}{c^{2}}
(\frac{3+2\omega_{BD}}{3+2\omega_{BD}+e^{-m_{0}r}})-2\Phi-2\psi,\\\nonumber
g_{0i}&\approx&\frac{1}{c^{3}}\left(4U_{i}
\left(\frac{3+2\omega_{BD}}{3+2\omega_{BD}+e^{-m_{0}r}}\right)
-\frac{1}{2}\frac{\partial^{2}\chi}{\partial t\partial
x_{i}}\right),\\\nonumber g_{ij}&\approx&[-1-\frac{2\gamma_{BD}
U}{c^{2}}-\frac{\Lambda_{BD}r^{2}}{3c^{2}}]\delta_{ij}.
\end{eqnarray}
The resulting Christoffel symbols are given in Appendix \textbf{A}.
The time component of Eq.(\ref{9}) gives
\begin{eqnarray}\nonumber
\frac{\partial T^{00}}{\partial x_{0}}+\frac{\partial
T^{0i}}{\partial x_{i}}+\left(\Gamma^{0}_{00}+z_{0}\right)T^{00}
+\left(2\Gamma^{0}_{0i}+z_{i}\right)T^{0i}+\Gamma^{0}_{ij}T^{ij}=0,
\end{eqnarray}
which yields
\begin{eqnarray}\nonumber
&&\frac{\partial\eta}{\partial t}+\frac{\partial \eta
v_{i}}{\partial
x_{i}}-\frac{\rho}{c^{2}}\left(\frac{3\exp^{-m_{0}}r}{3+2\omega_{BD}+e^{-m_{0}r}}\frac{\partial
U}{\partial t}\right.\\\nonumber
&&\left.+v_{i}\frac{\partial}{\partial
x_{i}}(\frac{3e^{-m_{0}r}}{3+2\omega_{BD}+e^{-m_{0}r}}U)+\frac{\Lambda_{BD}
r^{2}}{3}\right) +\frac{1}{c^{2}}\left(\rho\frac{\partial
U}{\partial t}-\frac{\partial p}{\partial t}\right)\\\label{10}
&&+v_{i}\frac{\partial}{\partial
x_{i}}\left(\frac{e^{-m_{0}r}}{3+2\omega_{BD}+e^{-m_{0}r}}U)+\frac{\Lambda_{BD}
r^{2}}{3}\right)=0,
\end{eqnarray}
where
\begin{equation}\nonumber
\eta=\rho \left[1+\frac{1}{c^2}(v^{2}+2U-\frac{2\Lambda_{BD}r^2}{3}
+\pi+\frac{p}{\rho})\right].
\end{equation}
This equation can replace the equation of continuity of Newtonian
gravity in MBD hydrodynamics. The spatial components of Eq.(\ref{9})
become
\begin{eqnarray}\nonumber
\frac{1}{c}\frac{\partial T{0i}}{\partial t}+\frac{\partial
T^{ij}}{\partial
x_{j}}+\Gamma^{i}_{00}T^{00}+2\Gamma^{i}_{j0}T^{0j}+z_{0}T^{i0}
+\Gamma^{i}_{jk}T^{jk}+z_{j}T^{ij}=0,
\end{eqnarray}
which can be written as
\begin{eqnarray}\nonumber
&&\frac{\partial \eta v_{i}}{\partial t}+\frac{\partial\eta
v_{i}v_{j}}{\partial x_{i}}+\frac{\partial}{\partial
x_{i}}\left[\left(1+2\gamma_{BD}U+\frac{\Lambda_{BD}
r^2}{3}\right)p\right]+\frac{2\rho}{c^{2}}\frac{d}{dt}
\left[\left(2\gamma_{BD}U\right.\right.\\\nonumber
&&\left.\left.+\frac{\Lambda_{BD}r^2}{3}\right)v_{i}\right]
-\frac{4\rho}{c^2}\frac{d}{dt}
\left[\frac{3+2\omega_{BD}}{3+2\omega_{BD}+e^{-m_{0}r}}U_{i}\right]\\\nonumber
&&-\frac{\rho}{c^2}\left[\frac{(3+2\omega_{BD}+e^{-m_{0}r})\sigma}{3+2\omega_{BD}}
\frac{\partial }{\partial
x_{i}}\left(\frac{3+2\omega_{BD}}{3+2\omega_{BD}+e^{-m_{0}r}}U\right)+\frac{\partial
\Phi'}{\partial x_{i}}\right]\\\nonumber
&&-\frac{4\rho}{c^{2}}\frac{\partial }{\partial
x_{j}}\left(\frac{3+2\omega_{BD}}{3+2\omega_{BD}+e^{-m_{0}r}}U_{i}\right)
-\frac{\rho}{c^2}\frac{\partial }{\partial
x_{i}}\left(\frac{3+2\omega_{BD}}{3+2\omega_{BD}+e^{-m_{0}r}}U\right)\\\label{11}
&&+\frac{\rho}{2c^{2}}\left(U_{i}-U_{\alpha;i\alpha}\right)-\frac{\rho}{2c^{2}}
W_{i}+\frac{\rho}{2c^{2}}Z_{i(BD)}=0,
\end{eqnarray}
where the potential functions $U_{\alpha;i\alpha},~ W_{i}$ and $Z_{i
(BD)}$ are described in Appendix \textbf{A}. Equations (\ref{10})
and (\ref{11}) provide generalized form of equation of continuity
and Euler equation of Newtonian hydrodynamics which represent
equation of motion in MBD gravity.

\subsection{Conservation Laws}

Here we discuss hydrodynamics of fluid by evaluating the basic
conservation laws in pN limits of MBD gravity \cite{2, 10}.

\subsubsection{The Conservation of Mass}

The conservation of mass implies that mass neither created nor
destroyed. The conserved mass can be calculated by integrating
Eq.(\ref{10}) over the volume occupied by the MBD fluid. The
resulting equation of continuity is given by
\begin{eqnarray}\label{12}
\frac{\partial \rho'}{\partial t}+\frac{\partial}{\partial
x_{i}}\left(\rho'v_{i}\right)=0,
\end{eqnarray}
where
\begin{equation}\nonumber
\rho'=\rho\left[1+\frac{1}{c^{2}}\left[\frac{1}{2}v^{2}-\frac{\Lambda_{BD}r^{2}}{3}
+\frac{9+6\omega_{BD}-e^{-m_{0}r}}{3+2\omega_{BD}+e^{-m_{0}r}}U\right]\right].
\end{equation}
This equation shows that the mass function expressed in terms of
density $\rho'$ remains conserved.

\subsubsection{The Conservation of Momentum}

The conservation equations of momentum help us to discuss dynamics
of fluid motion. The conservation law of linear momentum states that
the rate of change of total linear momentum is zero, i.e., the total
linear momentum of the system remains constant. For conserved linear
momentum, we integrate Eq.(\ref{11}) over the volume along with
boundary condition $p=0$ so that the conservation equation of total
linear momentum is
\begin{equation}\nonumber
\frac{d}{dt}\int_{v'}L_{i}dx=0,
\end{equation}
or
\begin{equation}\nonumber
\int_{v'}L_{i}dx=constant,
\end{equation}
where $L_{i}$ is the total linear momentum per unit volume given by
\begin{eqnarray}\nonumber
L_{i}&=&\eta
v_{i}+\frac{\rho}{2c^{2}}\left(U_{i}-U_{\alpha;i\alpha}\right)+\frac{2\rho}{c^{2}}
\left(\left(2U+\frac{\Lambda_{BD}
r^{2}}{3}\right)v_{i}\right.\\\label{13}
&-&\left.2(\frac{3+2\omega_{BD}}{3+2\omega_{BD}+e^{-m_{0}r}}U_{i})\right).
\end{eqnarray}
According to law of conservation of angular momentum in homogenous
space, the total angular momentum of the system remains constant.
This is evaluated by first multiplying the equation of motion by
$x_{j}$ and then subtracted it by the same expression with indices
$i$ and $j$ interchanged. The resultant is then integrated over the
volume $v'$ giving the equation of conserved total angular momentum
as
\begin{equation}\nonumber
\frac{d}{dt}\int_{v'}J_{ij}dx=0,
\end{equation}
or
\begin{equation}\nonumber
\int_{v'}J_{ij}=constant,
\end{equation}
with total angular momentum
\begin{equation}\label{14}
J_{ij}=x_{i}L_{j}-x_{j}L_{i}.
\end{equation}

\subsubsection{The Conservation of Energy}

The law of conservation of energy states that energy neither created
nor destroyed, i.e., total energy of the system remains constant. In
order to evaluate total energy of the system, we contract
Eq.(\ref{11}) with $v_{i}$ and then integrate over the volume $v'$.
Consequently, the conservation of energy is
\begin{equation}\nonumber
\frac{d}{dt}\int_{v'}\emph{E}dx=0,
\end{equation}
or
\begin{equation}\nonumber
\emph{E}=constant,
\end{equation}
where the total energy is given by
\begin{eqnarray}\nonumber
\emph{E}&=&\left(\eta-\frac{1}{2}\rho'\right)v^{2}+\rho'\Pi
-\frac{1}{2}\rho'\left(\frac{3+2\omega_{BD}U'}{3+2\omega_{BD}
+e^{-m_{0}r}}\right)
+\frac{1}{c^{2}}\rho\left(-\frac{v^{2}}{8}\right.\\\nonumber
&+&\left.\frac{1}{2}
\left(\frac{3+2\omega_{BD}U}{3+2\omega_{BD}+e^{-m_{0}r}}\right)^{2}
-\pi\left(\frac{3+2\omega_{BD}U}{3+2\omega_{BD}+e^{-m_{0}r}}\right)\right.\\\nonumber
&-&\left.\frac{1}{2}v^{2}\pi
-\frac{3}{2}\left(\frac{3+2\omega_{BD}v^{2}U}
{3+2\omega_{BD}+e^{-m_{0}r}}\right)+2v^{2}\left(2\gamma_{BD}U+\frac{\Lambda_{BD}
r^{2}}{3}\right)\right.\\\label{14a}
&+&\left.\frac{1+2\omega_{BD}+e^{-m_{0}r}}{4(3+\omega_{BD}+e^{-m_{0}r})}v_{i}U_{i}
-\frac{1}{4}v_{i}U_{\alpha;i\alpha}\right)+\frac{\rho_{BD}}{c^{2}}U_{i(BD)}.
\end{eqnarray}
Here, we assume super-potential function $U'$ as
\begin{eqnarray}\nonumber
&&\nabla^{2}\left(\frac{3+2\omega_{BD}}{3+2\omega_{BD}
+e^{-m_{0}r}}\right)U'=-4\pi
G_{eff}\left(\frac{3+2\omega_{BD}}{3+2\omega_{BD}
+e^{-m_{0}r}}\right)\rho'\\\nonumber &&=-4\pi
G_{eff}\left(\frac{3+2\omega_{BD}}{3+2\omega_{BD}
+e^{-m_{0}r}}\right)
\rho\left[1+\frac{1}{c^{2}}\left[\frac{1}{2}v^{2}-\frac{2\Lambda_{BD}r^{2}}{3}
\right.\right.\\\nonumber
&&\left.\left.+\frac{9+6\omega_{BD}}{3+2\omega_{BD}+e^{-m_{0}r}}U\right]\right].
\end{eqnarray}

\section{Stability in Massive Brans-Dicke Gravity}

The hydrostatic equilibrium is the state of fluid in which the
pressure gradient forces are balanced by all other forces (like
gravitational forces) and equation of motion does not depend upon
time ($v_{i}=0$). Thus, the hydrostatic conditions from
Eq.(\ref{11})-(\ref{14a}) are given by
\begin{eqnarray}\label{14'a}
\frac{\partial p}{\partial x_{i}}&=&\rho g_{BD},\\\label{14'b}
\rho'&=&\rho\left[1+\frac{1}{c^{2}}\left[-\frac{\Lambda_{BD}r^{2}}{3}
+\frac{9+6\omega_{BD}-e^{-m_{0}r}}{3+2\omega_{BD}+e^{-m_{0}r}}U\right]\right]
=Constant,\\\nonumber L_{i}&=&\eta
v_{i}+\frac{\rho}{2c^{2}}\left(U_{i}-U_{\alpha;i\alpha}\right)+\frac{2\rho}{c^{2}}
\left(\left(2U+\frac{\Lambda_{BD}
r^{2}}{3}\right)v_{i}\right.\\\label{14c}
&-&\left.2(\frac{3+2\omega_{BD}}{3+2\omega_{BD}
+e^{-m_{0}r}}U_{i})\right)=0.\\\label{14'c} J_{ij}&=&0,\\\nonumber
\emph{E}&=&
-\frac{1}{2}\rho'\left(\frac{3+2\omega_{BD}U'}{3+2\omega_{BD}
+e^{-m_{0}r}}\right) +\frac{1}{2c^{2}}\rho
\left(\frac{3+2\omega_{BD}U}{3+2\omega_{BD}+e^{-m_{0}r}}\right)^{2}\\\label{14'd}
&-&\pi\left(\frac{3+2\omega_{BD}U}{3
+2\omega_{BD}+e^{-m_{0}r}}\right).
\end{eqnarray}
with
\begin{eqnarray}\nonumber
g_{BD}&=&\left[\left(1+2\gamma_{BD}U+\frac{\Lambda_{BD}
r^2}{3}\right)\right]^{-1}\left[
\frac{1}{c^2}\left[\frac{(3+2\omega_{BD}+e^{-m_{0}r})\sigma}{3+2\omega_{BD}}
\right.\right.\\\nonumber &\times&\left.\left.\frac{\partial
}{\partial
x_{i}}\left(\frac{3+2\omega_{BD}}{3+2\omega_{BD}+e^{-m_{0}r}}U\right)+\frac{\partial
\Phi'}{\partial x_{i}}\right]-\frac{1}{c^2}\left[\frac{\partial
}{\partial x_{i}}\right.\right.\\\nonumber
&\times&\left.\left.\left(\frac{3+2\omega_{BD}}{3+2\omega_{BD}+e^{-m_{0}r}}U\right)
-\frac{p}{\rho}\frac{\partial}{\partial
x_{i}}\left[\left(1+2\gamma_{BD}U+\frac{\Lambda_{BD}
r^2}{3}\right)\right]\right]\right.\\\nonumber
&-&\left.\frac{1}{2c^{2}}Z'_{i(BD)}\right],
\end{eqnarray}
where the value of $Z'_{i(BD)}$ is given in Appendix \textbf{A}. The
term ``$\rho g_{BD}$'' represents total gravitational effects of MBD
systems in hydrostatic equilibrium. Equations
(\ref{14'a})-(\ref{14'd}) are stability conditions which describe
that in stable configuration (hydrostatic equilibrium) the total MBD
gravitational force balances the pressure gradient force due to
matter distribution. The total density $\rho'$ due to matter as well
as scalar field distribution becomes constant (the fluid apparently
becomes incompressible). The total linear as well as angular
momentum of the system vanishes.

The above discussion implies that in unstable configuration such as
gravitational collapse, the force of gravity takes over the pressure
force (hydrostatic equilibrium of the system is disturbed). The
total density of the system does not remain constant and momentum
induces into the system. The conditions of such type of instability
in MBD fluid can directly be obtained from
Eqs.(\ref{14'a})-(\ref{14'd}) as
\begin{eqnarray}\label{15'a}
\frac{\partial p}{\partial x_{i}}&<& \rho g_{BD},\\\label{15'b}
\rho'&=&\rho\left[1+\frac{1}{c^{2}}\left[-\frac{\Lambda_{BD}r^{2}}{3}
+\frac{9+6\omega_{BD}-e^{-m_{0}r}}{3+2\omega_{BD}+e^{-m_{0}r}}U\right]\right]
\neq Constant,\\\nonumber L_{i}&=&\eta
v_{i}+\frac{\rho}{2c^{2}}\left(U_{i}-U_{\alpha;i\alpha}\right)+\frac{2\rho}{c^{2}}
\left(\left(2U+\frac{\Lambda_{BD}
r^{2}}{3}\right)v_{i}\right.\\\nonumber
&-&\left.2(\frac{3+2\omega_{BD}}{3+2\omega_{BD}
+e^{-m_{0}r}}U_{i})\right)\neq0.\\\label{15'c}
J_{ij}&\neq&0,\\\nonumber \emph{E}&\neq&
-\frac{1}{2}\rho'\left(\frac{3+2\omega_{BD}U'}{3+2\omega_{BD}
+e^{-m_{0}r}}\right) +\frac{1}{2c^{2}}\rho
\left(\frac{3+2\omega_{BD}U}{3+2\omega_{BD}+e^{-m_{0}r}}\right)^{2}\\\label{15'd}
&-&\pi\left(\frac{3+2\omega_{BD}U}{3
+2\omega_{BD}+e^{-m_{0}r}}\right).
\end{eqnarray}

\section{Spherically Symmetric Barotropic Stars in Massive Brans-Dicke Gravity}

Barotropic is a state of fluid in which total density of a system is
a function of pressure only.  In homogenous and isotropic universe
(pN regimes), barotropic fluid satisfies the equation of state
\begin{equation}\label{16'}
p=w\rho,
\end{equation}
where $w$ is the equation of state parameter. Different values of
$w$ represent different cosmic configurations such as $w=0$ shows
dust, $w=1/3$ represents radiation era, $w=-1/3$ indicates cosmic
string, $w=-2/3$ expresses domain walls and $w<-1/3$ represents dark
energy dominated era. It is well-known that the configuration of
spherically symmetric stars depends upon time and radial components
only. Here, we represent dynamics of a spherical symmetric
barotropic stars in pN approximation of MBD gravity.

\subsection{Hydrodynamics}

Any spherically symmetric star in MBD gravity has the following
hydrodynamical quantities
\begin{itemize}
\item the conserved mass function is defined in terms of density
\begin{equation}\label{16}
\rho'=\rho\left[1+\frac{1}{c^{2}}\left[\frac{1}{2}v^{2}-\frac{\Lambda_{BD}r^{2}}{3}
+\frac{9+6\omega_{BD}-e^{-m_{0}r}}{3+2\omega_{BD}+e^{-m_{0}r}}U\right]\right],
\end{equation}
\item the total conserved linear momentum is given by
\begin{eqnarray}\nonumber
L_{r}&=&\eta
v_{r}+\frac{\rho}{2c^{2}}\left(U_{r}-U_{\alpha;r\alpha}\right)+\frac{2\rho}{c^{2}}
\left(\left(2U+\frac{\Lambda_{BD}
r^{2}}{3}\right)v_{r}\right.\\\label{17}
&-&\left.2(\frac{3+2\omega_{BD}}{3+2\omega_{BD}+e^{-m_{0}r}}U_{r})\right),
\end{eqnarray}
\item the non-zero components of conserved angular momentum are
\begin{eqnarray}\label{18}
J_{r\theta}=-\theta L_{r},\quad J_{r\phi'}=-\phi' L_{r},
\end{eqnarray}
\item the total conserved energy of the system is described as
\begin{eqnarray}\nonumber
\emph{E}&=&\left(\eta-\frac{1}{2}\rho'\right)v^{2}+\rho'\Pi
-\frac{1}{2}\rho'\left(\frac{3+2\omega_{BD}U'}{3+2\omega_{BD}
+e^{-m_{0}r}}\right)
+\frac{1}{c^{2}}\rho\left(-\frac{v^{2}}{8}\right.\\\nonumber
&+&\left.\frac{1}{2}
\left(\frac{3+2\omega_{BD}U}{3+2\omega_{BD}+e^{-m_{0}r}}\right)^{2}
-\pi\left(\frac{3+2\omega_{BD}U}{3+2\omega_{BD}+e^{-m_{0}r}}\right)\right.\\\nonumber
&-&\left.\frac{1}{2}v^{2}\pi
-\frac{3}{2}\left(\frac{3+2\omega_{BD}v^{2}U}
{3+2\omega_{BD}+e^{-m_{0}r}}\right)+2v^{2}\left(2\gamma_{BD}U+\frac{\Lambda_{BD}
r^{2}}{3}\right)\right.\\\label{19}
&+&\left.\frac{1+2\omega_{BD}+e^{-m_{0}r}}{4(3+\omega_{BD}+e^{-m_{0}r})}v_{r}U_{r}
-\frac{1}{4}v_{r}U_{\alpha;r\alpha}\right)+\frac{\rho_{BD}}{c^{2}}U_{r(BD)}.
\end{eqnarray}
\end{itemize}

\subsection{Conditions of Stability and Instability}

Any star remains stable as long as its hydrostatic equilibrium
remains stable. In hydrostatic equilibrium, a spherically symmetric
barotropic star satisfies the following conditions
\begin{equation}\label{20}
\frac{dp}{dr}=\rho g'_{BD}
\end{equation}
with
\begin{eqnarray}\nonumber
&&g'_{BD}=\left[1+2\gamma_{BD}U+\frac{\Lambda_{BD}
r^2}{3}\right]^{-1}\left[
\frac{1}{c^2}\left[\frac{(3+2\omega_{BD}+e^{-m_{0}r})\sigma}{3+2\omega_{BD}}
\right.\right.\\\nonumber
&&\left.\left.\times\frac{d}{dr}\left(\frac{3+2\omega_{BD}}{3+2\omega_{BD}+e^{-m_{0}r}}U\right)+\frac{d
\Phi'}{dr}\right] -\frac{1}{c^2}\frac{d}{dr}
\left(\frac{3+2\omega_{BD}}{3+2\omega_{BD}+e^{-m_{0}r}}U\right)\right.\\\nonumber
&&\left.-p\frac{d}{dr}\left[1+2\gamma_{BD}U+\frac{\Lambda_{BD}
r^2}{3}\right]-\frac{1}{2c^{2}}Z'_{r(BD)}\right].
\end{eqnarray}
\begin{eqnarray}\nonumber
\rho'&=&\rho\left[1+\frac{1}{c^{2}}\left[-\frac{\Lambda_{BD}r^{2}}{3}
+\frac{9+6\omega_{BD}-e^{-m_{0}r}}{3+2\omega_{BD}+e^{-m_{0}r}}U\right]\right]
=Constant\\\label{20'}\\\label{20''}
L_{r}&=&J_{r\theta}=J_{r\phi'}=0.
\end{eqnarray}
From Eqs.(\ref{16'}), (\ref{20}) and (\ref{20'}) we get
\begin{eqnarray}\label{20'a}
\rho&=&e^{(\frac{1}{w}\int g''_{BD}dr)},\\\label{20'b}
p&=&we^{(\frac{1}{w}\int g''_{BD}dr)},\\\nonumber
\rho'&=&e^{(\frac{1}{w}\int
g''_{BD}dr)}\left[1+\frac{1}{c^{2}}\left[-\frac{\Lambda_{BD}r^{2}}{3}
+\frac{9+6\omega_{BD}-e^{-m_{0}r}}{3+2\omega_{BD}+e^{-m_{0}r}}U\right]\right]
=Constant,\\\label{20'c}
\end{eqnarray}
where
\begin{eqnarray}\nonumber
&&g''_{BD}=\left[1+2\gamma_{BD}U+\frac{\Lambda_{BD}
r^2}{3}\right]^{-1}\left[
\frac{1}{c^2}\left[\frac{(3+2\omega_{BD}+e^{-m_{0}r})\sigma}{3+2\omega_{BD}}
\right.\right.\\\nonumber
&&\left.\left.\times\frac{d}{dr}\left(\frac{3+2\omega_{BD}}{3+2\omega_{BD}+e^{-m_{0}r}}U\right)+\frac{d
\Phi'}{dr}\right] -\frac{1}{c^2}\frac{d}{dr}
\left(\frac{3+2\omega_{BD}}{3+2\omega_{BD}+e^{-m_{0}r}}U\right)\right.\\\nonumber
&&\left.-w\frac{d}{dr}\left[1+2\gamma_{BD}U+\frac{\Lambda_{BD}
r^2}{3}\right]-\frac{1}{2c^{2}}Z''_{r(BD)}\right].
\end{eqnarray}
The value of $Z''_{r(BD)}$ is given in Appendix \textbf{A}.

Since in the phenomenon of gravitational collapse pressure gradient
forces are over take by gravitational forces. A barotropic star
becomes unstable (collapses) whenever
\begin{equation}\nonumber
w\frac{d\rho}{dr}<\rho g''_{BD}
\end{equation}
which implies
\begin{eqnarray}\label{21'a}
\rho&<&e^{(\frac{1}{w}\int g''_{BD}dr)},\\\label{21'b}
p&<&we^{(\frac{1}{w}\int g''_{BD}dr)},\\\label{21'c}
\rho'&<&e^{(\frac{1}{w}\int
g''_{BD}dr)}\left[1+\frac{1}{c^{2}}\left[-\frac{\Lambda_{BD}r^{2}}{3}
+\frac{9+6\omega_{BD}-e^{-m_{0}r}}{3+2\omega_{BD}+e^{-m_{0}r}}U\right]\right].
\end{eqnarray}

\subsection{Cosmic Strings}

Cosmic strings are $1$-dimensional topological defects that are
related to solitonic solutions of the classical equations for
complex scalar field. These types of matter satisfy equation of
state $p=-1/3\rho$ \cite{12}. They are considered to have immense
density and are significant source of gravitational waves. The
hydrodynamics of this kind of fluid in MD gravity are described by
Eqs.(\ref{16})-(\ref{19}) along with defined equation of state. The
conditions for instability configuration (gravitational collapse)
are
\begin{eqnarray}\nonumber
\rho&<&e^{(-3\int g''_{BD}dr)},\\\nonumber
p&<&\frac{-1}{3}e^{(-3\int g''_{BD}dr)},\\\nonumber
\rho'&<&e^{(-3\int
g''_{BD}dr)}\left[1+\frac{1}{c^{2}}\left[-\frac{\Lambda_{BD}r^{2}}{3}
+\frac{9+6\omega_{BD}-e^{-m_{0}r}}{3+2\omega_{BD}+e^{-m_{0}r}}U\right]\right],\\\nonumber
L_{r}&\neq&J_{r\theta}\neq J_{r\phi'}\neq0.
\end{eqnarray}

\subsection{Domain Wall}

Domain walls are $2$-dimensional topological defects in different
scalar fields. They are topological solitons which are considered as
a resultant of spontaneous broken of discrete symmetry. These are
perfect fluid types which obey equation of state $p=-2/3\rho$
\cite{13}. It is believed that collision of two such walls violently
emit gravitational waves. The equation of state along with
Eqs.(\ref{16})-(\ref{19}) represent hydrodynamics of domain wall in
pN regimes under the influence of MBD gravity. The respective
conditions for unstable configuration are
\begin{eqnarray}\nonumber
\rho&<&e^{(-\frac{3}{2}\int g''_{BD}dr)},\\\nonumber
p&<&\frac{-2}{3}e^{(-\frac{3}{2}\int g''_{BD}dr)},\\\nonumber
\rho'&<&e^{(-\frac{3}{2}\int
g''_{BD}dr)}\left[1+\frac{1}{c^{2}}\left[-\frac{\Lambda_{BD}r^{2}}{3}
+\frac{9+6\omega_{BD}-e^{-m_{0}r}}{3+2\omega_{BD}+e^{-m_{0}r}}U\right]\right],\\\nonumber
L_{r}&\neq&J_{r\theta}\neq J_{r\phi'}\neq0.
\end{eqnarray}

\subsection{Dust Fluid}

Dust is a state of fluid in which fluid particles are either
approximately stationary or moving in non-intersecting geodesics
producing zero pressure. Equation (\ref{16})-(\ref{19}) along with
$p=0$ describe hydrodynamics of dust in spherically symmetric
configuration. In this case the condition of stability (hydrostatic
equilibrium) can be directly obtained from Eq.(\ref{20}) as follows
\begin{equation}\nonumber
\rho g''_{BD}=0,
\end{equation}
since $\rho\neq0$, this implies $g''_{BD}=0$ which yield
\begin{eqnarray}\nonumber
Z'''_{r(BD)}&=&
\left[\frac{(3+2\omega_{BD}+e^{-m_{0}r})\sigma}{3+2\omega_{BD}}
\frac{d}{dr}\left(\frac{3+2\omega_{BD}}{3+2\omega_{BD}+e^{-m_{0}r}}U\right)+\frac{d
\Phi'}{dr}\right]\\\nonumber &-&\frac{d}{dr}
\left(\frac{3+2\omega_{BD}}{3+2\omega_{BD}+e^{-m_{0}r}}U\right),
\end{eqnarray}
where the value of $Z^{'''}_{r(BD)}$ is given in Appendix
\textbf{A}.

\section{Conclusions}

This paper is devoted to investigate hydrodynamics as well as
hydrostatic of MBD fluid in pN regime. For this purpose, we have
used simple perturbation scheme and evaluated complete pN
approximation of the field equations solution in terms of potential
functions of celestial masses. We have approximated $g_{00}$ to
$O(c^{-4}),~g_{0i}$ upto $O(c^{-3}),~g_{ij}$ to $O(c^{-2})$ and
perturbed scalar field upto $O(c^{-2})$. We have generalized
standard Newtonian equation of continuity and Euler equation of
motion in pN limits of MBD gravity. The equations governing
hydrodynamics, stability and instability of the fluid in weak-field
regimes are developed. These developed models are then used to
discuss stability conditions of a spherically symmetric star in pN
approximation of MBD gravity. In particular, we have discussed
conditions for unstable configuration of barotropic systems such as
dust, cosmic string and domain walls.

The obtained results provide a range of deviation of MBD theory from
BD and GR gravities in pN approximation. The governing equations
involve some extra and generalized potential functions that are not
used by BD and GR theories, which implies that celestial objects
used in MBD theory are more massive than those described by GR and
BD gravities \cite{2,10}. It is also found that MBD gravity reduces
to GR theory by freezing the dynamics at $\omega_{BD}\rightarrow
\infty, V_{0}\ll1$ and $\mid\frac{\varphi^{(2)}}{\phi_{0}}\mid
\ll1$. The derived model provides an essential (boot-strap)
character of modified gravity (MBD gravity) which can investigate
effects of dark energy on the hydrodynamics behavior of large-scale
systems or we can explore large-scale systems in accelerated
expanding universe.

\section*{Appendix A}

The contravariant and covariant components of four-velocity and
energy-momentum tensor are
\begin{eqnarray}\nonumber
u^{0}&=&1+\frac{1}{c^{2}}(\frac{1}{2}v^{2}+2U
-\frac{\Lambda_{BD}r^{2}}{3})+O(4),\\\nonumber
u_{0}&=&1+\frac{1}{c^{2}}(\frac{1}{2}v^{2}-2U
+\frac{\Lambda_{BD}r^2}{3})+O(4),\\\nonumber
u^{i}&=&\left[1+\frac{1}{c^{2}}(\frac{1}{2}v^{2}+2U
-\frac{\Lambda_{BD}r^{2}}{3})\right]\frac{v_{i}}{c}+O(4),\\\nonumber
\\\nonumber
u_{i}&=&-\frac{v_{i}}{c}+O(3),\\\nonumber T_{00}&=&\rho
c^{2}\left[1+\frac{1}{c^{2}}(v^{2}
-2U+\frac{2\Lambda_{BD}r^2}{3}+\pi)\right]+O(2),\\\nonumber
T^{00}&=&\rho c^{2}\left[1+\frac{1}{c^2}(v^{2})+2U
-\frac{2\Lambda_{BD}r^2}{3}+\pi\right]+O(2),\\\nonumber
T_{0i}&=&-\rho c v_{i}+O(1),
\\\nonumber
T^{0i}&=&\rho c
\left[1+\frac{1}{c^2}(v^{2}+2U-\frac{2\Lambda_{BD}r^2}{3}
+\pi+\frac{p}{\rho})\right]v_{i}+O(3),\\\nonumber T_{ij}&=&\rho
v^{i}v^{j}+\delta_{ij}p+O(2), \\\nonumber T^{ij}&=&\rho
v^{i}v^{j}+p\delta_{ij}+\frac{1}{c^{2}}
\left[\rho(v^{2}+2U-\frac{2\Lambda_{BD}r^{2}}{3}+\pi
+\frac{p}{\rho})v_{i}v_{j}\right.\\\nonumber
&-&\left.2p\gamma_{BD}U\delta_{ij}
-\frac{2\Lambda_{BD}r^2}{3}\delta_{ij}p\right]+O(4).
\end{eqnarray}
The Christoffel symbols in pN corrections are given by
\begin{eqnarray}\nonumber
\Gamma^{0}_{00}&=&-\frac{1}{c^{3}}
(\frac{3+2\omega_{BD}}{3+2\omega_{BD}+e^{-m_{0}r}})\frac{\partial
U}{\partial t},\\\nonumber
\Gamma^{0}_{0i}&=&-\frac{1}{c^{2}}\frac{\partial}{\partial
x_{i}}(\frac{3+2\omega_{BD}}{3+2\omega_{BD}+e^{-m_{0}r}}U),\\\nonumber
\Gamma^{0}_{ij}&=&\frac{1}{2c^{3}}\left[2\delta_{ij}\frac{\partial
\gamma_{BD}U}{\partial t}+4\left((\frac{\partial}{\partial
x_{j}})(\frac{3+2\omega_{BD}}{3+2\omega_{BD}+e^{-m_{0}r}}
U_{i})\right.\right.\\\nonumber
&+&\left.\left.(\frac{\partial}{\partial
x_{i}})(\frac{3+2\omega_{BD}}{3+2\omega_{BD}+e^{-m_{0}r}}
U_{j})\right)-\frac{\partial^{2}\chi}{\partial t \partial x_{i}
\partial x{j}}\right], \\\nonumber
\Gamma^{i}_{00}&=&-\frac{1}{c^{2}}\frac{\partial}{\partial x_{i}}
(\frac{3+2\omega_{BD}}{3+2\omega_{BD}+e^{-m_{0}r}}U)
+\frac{1}{c^{4}}\left[-\frac{\partial}{\partial
x_{i}}\left(\frac{1}{2}(h^{(2)}_{00})^{2}\right.\right.\\\nonumber
&-&\left.\left.\frac{1}{2}
(\frac{\varphi^{(2)}}{\phi_{0}})^2-\Phi-\psi\right)
-4\frac{3+2\omega_{BD}}{3+2\omega_{BD}+e^{-m_{0}r}}\frac{\partial
U_{i}}{\partial t}+\frac{1}{2}\frac{\partial^{2}\chi}{\partial
t\partial x_{i}}\right],\\\nonumber \Gamma^{i}_{0j}&=&
\frac{1}{c^{3}}\left[\gamma_{BD}\frac{\partial U}{\partial
t}\delta_{ij}-2\left((\frac{\partial}{\partial
x_{j}})(\frac{3+2\omega_{BD}}{3+2\omega_{BD}+e^{-m_{0}r}}
U_{i})\right.\right.\\\nonumber
&+&\left.\left.(\frac{\partial}{\partial
x_{i}})(\frac{3+2\omega_{BD}}{3+2\omega_{BD}+e^{-m_{0}r}}
U_{j})\right)\right],\\\nonumber
\Gamma^{i}_{jk}&=&\frac{1}{2c^{2}}\left[\frac{\partial}{\partial
x_{k}}(2\gamma_{BD}+\frac{\Lambda_{BD}r^2}{3})\delta_{ij}+\frac{\partial}{\partial
x_{j}}(2\gamma_{BD}+\frac{\Lambda_{BD}r^2}{3})\delta_{ik}\right.\\\nonumber
&-&\left.\frac{\partial}{\partial
x_{i}}(2\gamma_{BD}+\frac{\Lambda_{BD}r^2}{3})\delta_{jk}\right],
\end{eqnarray}
and
\begin{eqnarray}\nonumber
\Gamma^{\mu}_{\alpha\mu}=z_{\alpha}=\frac{\partial (log
\sqrt{-g})}{\partial
x_{\alpha}}=\frac{2}{c^{2}}\frac{\partial}{\partial
x_{\alpha}}\left[2U\gamma_{BD}+\frac{\Lambda_{BD}r^{2}}{3}\right],
\end{eqnarray}
which gives
\begin{eqnarray}\label{a}
\Gamma^{\mu}_{0\mu}=z_{0}=\frac{2}{c^{3}}\frac{\partial
(\gamma_{BD}U)}{\partial t},\quad
\Gamma^{\mu}_{i\mu}=z_{i}=\frac{\partial}{\partial
x_{i}}\left[2\gamma_{BD}U+\frac{\Lambda_{BD}r^2}{3}\right].
\end{eqnarray}
The potential functions $U_{\alpha;i\alpha},~W_{i}(\textbf{x})$ and
$Z_{i(BD)}$ expressed in generalized Euler equation of motion are
defined by
\begin{eqnarray}\nonumber
U_{\alpha;i\alpha}&=&G_{eff}\int_{v}\rho(\textbf{x}')v_{\alpha}(\textbf{x}')
\frac{(x_{i}-x'_{i})(x_{i}-x'_{i})dx'}{\mid
\textbf{x}-\textbf{x'}\mid^{3}},\\\nonumber
W_{i}(\textbf{x})&=&v_{\alpha}\frac{\partial}{\partial
x_{\alpha}}\left(U_{i}-U_{j;ij}\right)=
-G_{eff}\int_{v}\rho({\textbf{x}'})v_{i}({\textbf{x}})v_{i}({\textbf{x}'})
\frac{(x_{i}-x'_{i})dx'}{\mid
\textbf{x}-\textbf{x}'\mid^{3}}\\\nonumber
&-&G_{eff}\int_{v}\rho(x')
\left[v_{i}(\textbf{x})v_{\alpha}({\textbf{x}'})+v_{i}({\textbf{x}'})v_{\alpha}(\textbf{x})\right]
\frac{(x_{\alpha}-x'_{\alpha})dx'}{\mid
\textbf{x}-\textbf{x}'\mid^{3}}\\\nonumber
&+&3G_{eff}\int_{v}\rho({\textbf{x}'})\left[v_{\alpha}({\textbf{x}})
v_{\beta}(\textbf{x}')(x_{\alpha}-x'_{\alpha})(x_{\beta}-x'_{\beta})\right]
\frac{x_{i}-x'_{i}}{\mid\textbf{x}-\textbf{x}'\mid^{5}} ,\\\nonumber
\frac{\rho}{2c^{2}}Z_{i(BD)}&=&\frac{\rho_{BD}}{c^{2}}\frac{\partial
U_{BD}}{\partial x_{i}}=\rho\left[-2\left(1+\frac{e^{-m_{0}r}}
{3+2\omega_{BD}+e^{-m_{0}r}}U\right)\right.\\\nonumber
&+&\left.\frac{\Lambda_{BD} r^{2}}{3}+\frac{2\rho
v^{2}}{c^{2}}\left(\frac{\partial U}{\partial
x_{i}}+\frac{\partial}{\partial
x_{i}}(\frac{e^{-m_{0}r}}{3+2\omega_{BD}+e^{-m_{0}r}}U)\right.\right.\\\nonumber
&+&\left.\left.\frac{\partial(\Lambda_{BD}U)}{\partial
x_{i}}+\frac{\partial \Lambda_{BD}r^{2}}{\partial
x_{i}}\right)+\frac{p}{\rho c^{2}}\left(2\frac{\partial U}{\partial
x_{i}}\right.\right.\\\nonumber
&+&\left.\left.+2\frac{\partial}{\partial
x_{i}}(\frac{e^{-m_{0}r}}{3+2\omega_{BD}}U)
+2\frac{\partial(\Lambda_{BD}U)}{\partial
x_{i}}+\frac{1}{2}\frac{\partial \Lambda_{BD}r^{2}}{\partial
x_{i}}\right)\right.\\\nonumber &+&\left.\left(U-\frac{\Lambda_{BD}
r^{2}}{6}\right)\frac{\partial \Lambda_{BD} r^{2}}{\partial
x_{i}}\right].
\end{eqnarray}
The values of $Z'_{i(BD)}$, $Z''_{r(BD)}$ and $Z^{'''}_{r(BD)}$are
given by
\begin{eqnarray}\nonumber
\frac{\rho}{2c^{2}}Z'_{i(BD)}&=&\frac{\rho'_{BD}}{c^{2}}\frac{\partial
U_{BD}}{\partial x_{i}}=\rho\left[-\frac{2}{c^{2}}\left(1
+\frac{e^{-m_{0}r}}{3+2\omega_{BD}+e^{-m_{0}r}}U\right)
+\frac{\Lambda_{BD}
r^{2}}{3c^{2}}\right.\\\nonumber&+&\left.\frac{p}{\rho
c^{2}}\left(2\frac{\partial U}{\partial
x_{i}}+2\frac{\partial}{\partial
x_{i}}(\frac{e^{-m_{0}r}}{3+2\omega_{BD}+e^{-m_{0}r}}U)
+2\frac{\partial(\gamma_{BD}U)}{\partial
x_{i}}\right.\right.\\\nonumber
&+&\left.\left.\frac{1}{2}\frac{\partial \Lambda_{BD}r^{2}}{\partial
x_{i}}\right)+\left(U-\frac{\Lambda_{BD}
r^{2}}{6}\right)\frac{\partial \Lambda_{BD}r^{2}}{\partial
x_{i}}\right],\\\nonumber
\frac{\rho}{2c^{2}}Z''_{r(BD)}&=&\frac{\rho'_{BD}}{c^{2}}\frac{\partial
U_{BD}}{\partial x_{i}}=\rho\left[-\frac{2}{c^{2}}
\left(1+\frac{e^{-m_{0}r}}{3+2\omega_{BD}+e^{-m_{0}r}}U\right)
+\frac{\Lambda_{BD}
r^{2}}{3c^{2}}\right.\\\nonumber&+&\left.\frac{w}{c^{2}}\left(2\frac{d
U}{dr}+2\frac{d}{dr}(\frac{e^{-m_{0}r}}{3+2\omega_{BD}+e^{-m_{0}r}}U)
+2\frac{d(\gamma_{BD}U)}{dr}\right.\right.\\\nonumber
&+&\left.\left.\frac{1}{2}\frac{d
\Lambda_{BD}r^{2}}{dr}\right)+\left(U-\frac{\Lambda_{BD}
r^{2}}{6}\right)\frac{d\Lambda_{BD} r^{2}}{dr}\right],\\\nonumber
Z'''_{r(BD)}&=&\left[-\frac{2}{c^{2}}
\left(1+\frac{e^{-m_{0}r}}{3+2\omega_{BD}+e^{-m_{0}r}}U\right)
+\frac{\Lambda_{BD}r^{2}}{3c^{2}}\right.\\\nonumber
&+&\left.\left(U-\frac{\Lambda_{BD}
r^{2}}{6}\right)\frac{d\Lambda_{BD} r^{2}}{dr}\right].
\end{eqnarray}


\begin{thebibliography}{40}

\bibitem{1} Wilson, J.R. and
Mathews, G.J.: \emph{Relativistic Numerical Hydrodynamics}
(Cambridge University Press, 2007); Rezollz, L. and Zonotti, O.:
\emph{Relativistic Astrophysics} (Oxford University Press, 2013).

\bibitem{2*} Ayal, S. et al.: Astrophys. J. \textbf{550}(2001)846; Marek, A. et al.: Astron. Astrophys.
\textbf{445}(2006)273; Straumann, N.: \emph{General Relativity with
Applications to Astrophysics} (Springer, 2013).

\bibitem{2} Chanrasekhar, S.: Astrophys. J. \textbf{142}(1964)1488.

\bibitem{3} Chandareskhar, S.: \emph{Relativistic Astrophysics} (University of Chicago Press,
1990).
\bibitem{4} Herrera, L. et al.: Mon. Not. Roy. Astron. Soc. \textbf{237}(1989)257;
Chan, R. et al.: Mon. Not. Roy. Astron. Soc. \textbf{239}(1989)91;
Chan, R. and Herrera, L. et al.: Mon. Not. Roy. Astron. Soc.
\textbf{265}(1993)533; ibid. \textbf{267}(1994)637; Herrera, L. et
al.: Gen. Relativ. Gravit. \textbf{44}(2012)1143; Sharif, M. and
Azam, M.: J. Cosmol. Astropart. Phys. \textbf{02}(2012)043; Gen.
Relativ. Gravit. \textbf{44}(2012)1181; Mon. Not. Roy. Astron. Soc.
\textbf{430}(2013)3048.

\bibitem{5} Dirac, P.A.M.: Proc. R. Soc. Lond. A
\textbf{165}(1938)199; Brans, C.H. and Dicke, R.H.: Phys. Rev.
\textbf{124}(1961)925.

\bibitem{5a} Weinberg, E.J.: Phys. Rev. D \textbf{40}(1989)3950.

\bibitem{5b} Reasenberg, R.D. et al.: Astrophys. J.
\textbf{234}(1979)L219.

\bibitem{6} Bertotti, B.I.L. and Tortora, P.: Nature
\textbf{425}(2003)374; Felice, A.D. et al.: Phys. Rev. D
\textbf{74}(2006)103005.

\bibitem{6a} Santos, C. and Gregory, R.: Annals. Phys.
\textbf{258}(1997)111.

\bibitem{7} Banerjee, N. and Pavon, D.: Phys. Rev. D
\textbf{63}(2001)043504.

\bibitem{8} Mak, M.K. and Harko, T.: Europhys. Lett. \textbf{60}(2002)155;
 Bertolami, O. and Martins, P.J.: Phys. Rev. D
\textbf{61}(2000)064007; Bisaby, Y.: Astrophys. Space Sci.
\textbf{1}(2012)339; Sharif, M. and Waheed, S.: Eur. Phys. J. C
\textbf{72}(2012)1876; J. Phys. Soc. Jpn. \textbf{81}(2012)114901.

\bibitem{9} Perivolaropoulos, L.: Phys. Rev. D
\textbf{81}(2010)047501; Sharif, M. and Yousaf, Z.: Phys. Rev. D
\textbf{88}(2013)024020; Eur. Phys. J. C \textbf{73}(2013)2633; Mon.
Not. Roy. Astron. Soc. \textbf{432}(2013)264; ibid.
\textbf{434}(2013)2529; Sharif, M. and Manzoor, R.:  Mod. Phys.
Lett. A \textbf{29}(2014)1450192; Astrophys. Space Sci.
\textbf{354}(2014)497; ibid. \textbf{359}(2015)17; Phys. Rev. D
\textbf{91}(2015)024018; Gen. Relativ. Gravit. \textbf{47}(2015)98.

\bibitem{10} Nutku, Y.: Astrophys. J. \textbf{155}(1969)999.

\bibitem{11} Olmo, G.J.: Phys. Rev. D \textbf{72}(2005)083505.


\bibitem{12} Mukhanov, V.: \emph{Physical Foundations of Cosmology} (Cambridge University Press, 2005).

\bibitem{13} Weinberg, S.: \emph{The Quantum Theory of Fields} (Cambridge University Press,
1995).
\end{thebibliography}
\end{document}